\documentclass[twocolumn]{aastex631} 
\usepackage[utf8]{inputenc}
\usepackage{textgreek}
\usepackage{graphicx}
\graphicspath{ {./plots/} }
\usepackage{amsmath}
\usepackage{comment}
\usepackage{longtable}
\usepackage{booktabs}
\usepackage{threeparttable}
\usepackage{bm}

\newcommand\aastex{AAS\TeX}

\graphicspath{{./}{figures/}}
\begin{document}

\title{Template \aastex Article with Examples: 
v6.3.1\footnote{Released on March, 1st, 2021}}

\title{ODIN: Clustering Analysis of 14,000 Ly$\alpha$ Emitting Galaxies at z=2.4, 3.1, and 4.5}

\author[0000-0003-2986-8594]{Danisbel Herrera}
\affiliation{Department of Physics and Astronomy, Rutgers, the State University of New Jersey, Piscataway, NJ 08854, USA}

\author[0000-0003-1530-8713]{Eric Gawiser}
\affiliation{Department of Physics and Astronomy, Rutgers, the State University of New Jersey, Piscataway, NJ 08854, USA}
\affiliation{School of Natural Sciences, Institute for Advanced Study, Princeton, NJ 08540, USA}

\author{Barbara Benda}
\affiliation{Department of Physics, University of Washington,  Seattle, WA 98105, USA},

\author[0000-0002-9811-2443]{Nicole M. Firestone}
\affiliation{Department of Physics and Astronomy, Rutgers, the State University of New Jersey, Piscataway, NJ 08854, USA}
\author[0000-0002-9176-7252]{Vandana Ramakrishnan}
\affiliation{Department of Physics and Astronomy, Purdue University, 525 Northwestern Ave., West Lafayette, IN 47906, USA}

\author[0009-0008-4022-3870]{Byeongha Moon}
\affiliation{Korea Astronomy and Space Science Institute, 776 Daedeokdae-ro, Yuseong-gu, Daejeon 34055, Republic of Korea}

\author[0000-0003-3004-9596]{Kyoung-Soo Lee}
\affiliation{Department of Physics and Astronomy, Purdue University, 525 Northwestern Ave., West Lafayette, IN 47906, USA}

\author[0000-0001-9521-6397]{Changbom Park}
\affiliation{Korea Institute for Advanced Study, 85 Hoegi-ro, Dongdaemun-gu, Seoul 02455, Republic of Korea}

\author[0000-0001-5567-1301]{Francisco Valdes}
\affiliation{NSF’s National Optical-Infrared Astronomy Research Laboratory, 950 N. Cherry Ave., Tucson, AZ 85719, USA}

\author[0000-0003-3078-2763]{Yujin Yang}
\affiliation{Korea Astronomy and Space Science Institute, 776 Daedeokdae-ro, Yuseong-gu, Daejeon 34055, Republic of Korea}

\author[0000-0003-0570-785X]{Mar\'ia Celeste Artale}
\affiliation{Universidad Andres Bello, Facultad de Ciencias Exactas, Departamento de Fisica, Instituto de Astrofisica, Fernandez Concha 700, Las Condes, Santiago (RM), Chile}

\author[0000-0002-1328-0211]{Robin Ciardullo}
\affiliation{Department of Astronomy \& Astrophysics, The Pennsylvania
State University, University Park, PA 16802, USA}
\affiliation{Institute for Gravitation and the Cosmos, The Pennsylvania
State University, University Park, PA 16802, USA}

\author[0000-0001-6842-2371]{Caryl Gronwall}
\affiliation{Department of Astronomy \& Astrophysics, The Pennsylvania
State University, University Park, PA 16802, USA}
\affiliation{Institute for Gravitation and the Cosmos, The Pennsylvania
State University, University Park, PA 16802, USA}

\author[0000-0002-4902-0075]{Lucia Guaita}
\affiliation{Universidad Andres Bello, Facultad de Ciencias Exactas, Departamento de Fisica y Astronomia, Instituto de Astrofisica, Fernandez Concha 700, Las Condes, Santiago RM, Chile}
\affiliation{Millennium Nucleus for Galaxies}

\author[0000-0003-3428-7612]{Ho Seong Hwang}
\affiliation{Department of Physics and Astronomy, Seoul National University, 1 Gwanak-ro, Gwanak-gu, Seoul 08826, Republic of Korea}
\affiliation{SNU Astronomy Research Center, Seoul National University, 1 Gwanak-ro, Gwanak-gu, Seoul 08826, Republic of Korea}
\affiliation{Australian Astronomical Optics - Macquarie University, 105 Delhi Road, North Ryde, NSW 2113, Australia}

\author[0009-0001-4745-3555]{Jacob Kennedy}
\affiliation{Department of Physics and Astronomy, Rutgers, the State University of New Jersey, Piscataway, NJ 08854, USA}

\author[0000-0001-6270-3527]{Ankit Kumar}
\affiliation{Universidad Andres Bello, Facultad de Ciencias Exactas, Departamento de Fisica y Astronomia, Instituto de Astrofisica, Fernandez Concha 700, Las Condes, Santiago RM, Chile}

\author[0000-0001-6047-8469]{Ann Zabludoff}
\affiliation{Department of Astronomy, University of Arizona, 933 North Cherry Avenue, Rm. N204, Tucson, AZ 85721, USA}



\shorttitle{Clustering of Ly$\alpha$ Emitters} 
\shortauthors{Herrera et al.}

\begin{abstract}


Lyman Alpha Emitters (LAEs) are star-forming galaxies 
that efficiently probe the spatial distribution of galaxies in the high redshift universe. The spatial clustering of LAEs reflects the properties of their individual host dark matter halos, allowing us to study the evolution of the galaxy-halo connection. We analyze the clustering of 5233, 5220, and 3706 LAEs at $z$ = 2.4, 3.1, and 4.5, respectively, in the 9 deg$^2$ COSMOS field from the One-hundred-deg$^2$ DECam Imaging in Narrowbands (ODIN) survey. After correcting for redshift space distortions, LAE contamination rates, and the integral constraint, the observed angular correlation functions imply linear galaxy bias factors of $b$ = $1.72^{+0.26}_{-0.27}, 2.01^{+0.26}_{-0.29},$ and $2.95^{+0.40}_{-0.46}$, for $z$ = 2.4, 3.1, and 4.5, respectively. The median dark matter halo masses inferred from these measurements are $\log(M_{h}/M_{\odot})$ = $11.44^{+0.30}_{-0.28}, 11.13^{+0.26}_{-0.26}, \textnormal{ and } 10.85^{+0.24}_{-0.24}$ for the three samples, respectively. The analysis also reveals that LAEs occupy roughly 3-7\% of the halos whose clustering strength matches that of the LAEs. 

\end{abstract}


\section{Introduction} \label{sec:intro}

 Hierarchical structure formation predicts that galaxies are formed when gas cools and condenses within the potential wells of dark matter halos \citep{WhiteRees}. This theory indicates that there is a fundamental relationship between galaxies and their host halos that influences their evolution and spatial clustering. Studying the distribution of galaxies therefore allows us to probe the galaxy-halo connection to reveal key processes driving galaxy formation and evolution. 




We can study the large-scale features of a galaxy population by using a large, homogeneous sample of galaxies. Ly$\alpha$ emitters (LAEs) are predominantly young, nearly dust-free galaxies known to be undergoing a strong period of star-formation at the time of observation \citep{Hu_1996}.  These galaxies are characterized by their prominent Lyman-alpha (Ly$\alpha$) emission line, whose exceptionally strong observed-frame equivalent width allows us to obtain large observational samples with reliable photometric redshifts. 
Therefore LAEs are efficient probes of the spatial distribution of high-redshift galaxies, making them excellent candidates for a clustering analysis.

Clustering is defined as the excess probability of detecting a pair of galaxies relative to a randomly distributed pair with the same separation \citep{Peebles_2020}. We quantify the clustering strength of a galaxy sample through its correlation function, which allows us to determine the number of galaxy pairs in various spatial separation bins. A clustering analysis involves comparing this to the clustering of the underlying dark matter. 
By studying the LAE correlation function for various redshift slices of the universe, we can infer the evolution of galaxy and halo properties and their role in galaxy evolution. 

There have been several LAE clustering analyses reported in recent years. \cite{Kova__2007} used a sample of 151 LAEs at z $\approx$ 4.5 selected from the Large Area Lyman Alpha (LALA; \citealt{LALA}) survey and reported the galaxy bias, median host halo mass, and fraction of such dark matter halos that appear to host an LAE. 
\cite{Gawiser07} conducted a similar analysis of 162 LAEs at z $\simeq$ 3.1 from the MUSYC survey. 
\cite{Guaita10} used a slightly larger sample of 250 LAEs at z $\simeq$ 2.1 from the MUSYC survey.
\cite{Ouchi10} worked with a comparable sample of 207 LAEs at $z=$ 6.6 using Subaru data from the XMM-Newton Deep Survey field. This analysis reported bias values between 3--6 and average halo mass values of $10^{10}-10^{11}M_{\odot}$.
More recently, \cite{White} introduced an improved model to correct for redshift space distortions and analyzed the clustering of 4000 LAEs at $z = $2.4 and 3.1,  
with a significant reduction in uncertainties due to the larger sample sizes. 
After the SILVERRUSH program was introduced, \cite{Ouchi18} analyzed 2000 LAEs at $z =$ 5.7 and $z =$6.6. 
The most recent such analysis was conducted by \cite{umeda}, and includes a significantly larger sample of 20,000 LAEs across from the SILVERRUSH and CHORUS surveys. This set of galaxies comes from a wide field of 24 deg$^2$ and covers 6 redshifts slices from $z=$2.2 to 7.3, with the largest lower redshift sample containing 6995 LAEs and the highest redshift sample containing 5.   
The larger sample size 
at $z=$2.2 allowed clustering measurements with the smallest uncertainties thus far.

    To obtain increasingly precise measurements of LAEs and their halo properties, we need to pursue surveys with deeper imaging that cover more area. With this goal in mind, this paper reports a clustering analysis of over 14,000 LAEs  at $z=$2.4, 3.1, and 4.5. The LAE samples at $z=3.1$ and $z=4.5$ are the largest ever published at those redshifts, with 5220 and 3706 LAEs, respectively, with the $z=2.4$ sample of 5233 LAEs somewhat smaller than the $z=2.2$ sample from SILVERRUSH. 
Section~\ref{sec2} discusses LAE selection methods. Section \ref{sec3} describes the methodology used to perform the clustering analysis, including LAE and dark matter angular correlation function calculations and halo property derivations. We report results in \S\ref{sec4}, discuss their implications in \S\ref{sec5}, and conclude in \S\ref{conc}.  
We assume a $\Lambda$CDM model with cosmological parameters $\Omega_c = 0.25, \Omega_b = 0.048, h = 0.68, n_s = 0.97 \text{and} \sigma_8 = 0.8$ \citep{act}.

\section{LAE Samples}\label{sec2}
\subsection{The ODIN Survey}

The One-hundred-deg$^2$ DECam Imaging in Narrowbands (ODIN) is a narrowband survey that aims to study LAEs and Lyman-break galaxies and protoclusters, as well as their connection to filaments in the cosmic web \citep{ODIN,Ramakrishnan24}. The survey covers seven fields: COSMOS, Deep2-3, SHELA, XMM-LSS, CDF-S, EDF-S, and ELAIS-S1. 
The narrowband observations are acquired via the Dark Energy Camera (DECam) on the Blanco 4m telescope in Cerro-Tololo Inter-American Observatory in Chile. Three custom filters were built for this survey, $N419$, $N501$, and $N673$ which are sensitive to light with central wavelengths of 419, 501, and 673~nm, with full width at half maximum values of 7.2, 7.4, and 9.8~nm, respectively. 
These wavelengths were chosen because they correspond to the Ly$\alpha$ centered at redshifts of $z = $2.4, 3.1, and 4.5 with $\Delta z = $ 0.061, 0.062, and 0.082, respectively. 
The complete survey will cover a wide field of 100 deg$^2$, a large portion of which will overlap  LSST deep drilling fields. The high depth and wide area coverage should yield a final sample of over 100,000 LAEs. 

This work focuses on the COSMOS field, since it is the first field that ODIN  observed completely in all filters. The field originally covered 9 deg$^2$, but this was reduced to 7.53 deg$^2$  after applying a starmask to avoid bright stars. 

In addition to our narrowband imaging, we also used broadband imaging in the \textit{u} band from the CFHT Large Area U-band Deep Survey \citep{CFHT}, as well as in the \textit{g, r, i, z}, and \textit{y} bands from the Hyper
SuprimeCam Subaru Strategic Program data \citep{SubaruHSC}.
We then used Source Extractor \citep{SE} to create source catalogs \citep{ramakrishnan} from which we selected our LAE sample \citep{Nicole24}.  

\subsection{LAE Candidate Selection} 


Here we provide a brief overview of the selection criteria for our LAE candidates. For a more thorough breakdown of the LAE selection process, see \cite{Nicole24}.

From our Source Extractor (SE) 
 catalog we require 5 main criteria for an object to be considered an LAE.  First, we require that the signal-to-noise ratio in the narrow-band detection filter is $\geq$ 5, which ensures we have high quality object detection. 
 Second, the flux densities must be non-zero in the broad-band filters used to estimate narrow-band excess, in order to select objects with no missing data. Third, we require that the SE \texttt{FLAGS} parameter be $<$ 4 in order to ensure that we exclude objects with saturated pixels. Next, we perform a narrowband excess cut to exclude objects with rest frame Ly$\alpha$ equivalent width EW$_0 < 20$ \AA. 
Lastly, we require our LAE candidates to have a $(\text{BB} - \text{NB})$ color $\geq$ $3\sigma_{(\text{BB} - \text{NB})}$ to ensure a robust narrowband excess and reduce the risk of continuum-only contaminants. \footnote{As explained in \citet{Nicole24}, the narrowband excess color BB-NB is calculated using a hybrid-weighted double-broadband continuum estimation that uses a weighted average of the magnitudes in two nearby broadband filters to predict the continuum level at the narrowband wavelength.  For low signal-to-noise photometry, the same weights are used to average the flux densities instead of magnitudes.  $\sigma_{(\text{BB}-\text{NB})}$ is the uncertainty in this color calculated by propagating the individual photometric uncertainties in the narrowband and the two broadbands.} 

\section{LAE Clustering Methodology}\label{sec3}

\subsection{Angular Correlation Function}

The transmission curve of the filter allows us to extract a well-defined expected redshift distribution, but we cannot determine the specific redshift of individual sources. Redshifts are required for an analysis of the correlation function in real-space because they allow us to determine physical distances between sources. Since we do not have this 3D information, we measured the angular correlation function, $\omega(\theta)$. This is defined as the excess probability above the Poisson expectation of finding a pair of galaxies with an angular separation, $\theta$. To measure the correlation functions we use the Landy-Szalay (LS) estimator \citep{Landy_1993}:
\begin{equation} 
    \omega_{\text{LS}}(\theta)=\frac{DD(\theta)-2DR(\theta)+RR(\theta)}{RR(\theta)}
\end{equation}
where $DD(\theta)$ is the normalized pair counts at separation $\theta$ for the observed sample, $RR(\theta)$ is the normalized pair counts for the random sample, and $DR(\theta)$ is the normalized pair counts between the observed and random samples.
We use the python package \texttt{TreeCorr} \citep{Jarvis_2004} to obtain these pair counts.
We calculate the uncertainty intervals on $\omega_{LS}(\theta)$ using jacknife resampling with 16 patches.

The angular correlation function obtained from the LS estimator can be modeled as an intrinsic galaxy correlation function, $\omega_g(\theta)$, minus an integral constraint that arises from the LAE sky density estimation \citep{Peebles_2020}.
We define our model as

\begin{equation}\label{model1}
    1+\omega_{\text{LS}}(\theta)=\frac{1+\omega_g(\theta)}{1+\omega_{\Omega}}
\end{equation}
where the integral constraint is defined as
\begin{equation}\label{uglyIC}
    \omega_{\Omega}=\frac{1}{\Omega}\int \omega_g(\theta) d\Omega_1 d\Omega_2.
\end{equation} 
We rewrite equations \ref{model1} and \ref{uglyIC} in more useful terms as
\begin{equation}\label{convert_model}
    \omega_{\text{LS}}(\theta)=\frac{\omega_g(\theta)-\omega_{\Omega}}{1+\omega_{\Omega}}
\end{equation}
\begin{equation}\label{niceIC}
    \omega_{\Omega}=\frac{\sum RR(\theta) \omega_g(\theta)}{\sum RR(\theta)}
\end{equation}
and note that a random catalog will be needed both to measure the correlation function and to estimate the integral constraint.  

\subsection{Random Catalog}

\begin{figure*}[ht]
\includegraphics[width=\textwidth]{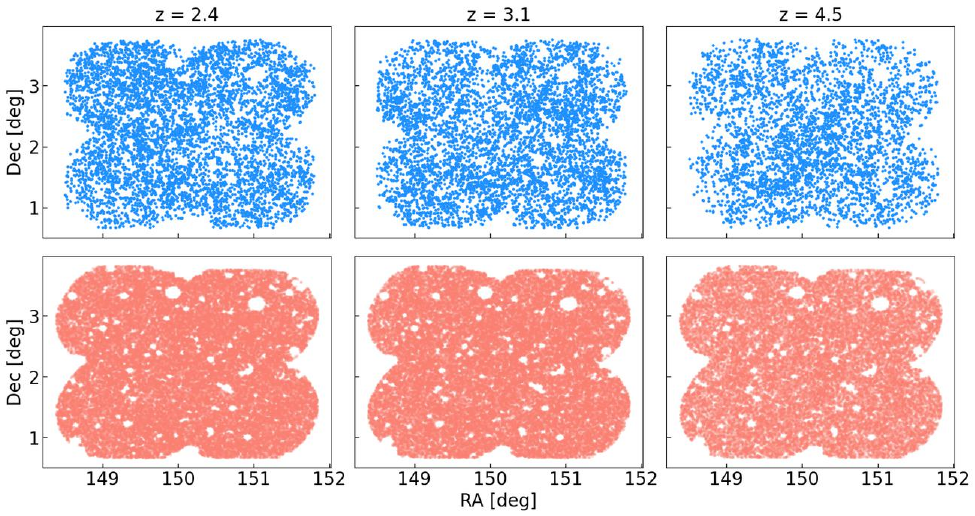}
\caption{The ODIN LAEs (top) and their corresponding random catalogs (bottom) in the COSMOS field. The random catalogs shown here have been sub-sampled to plot 10x as many objects as the LAE catalog at each redshift. 
}\label{fig:cat}
\end{figure*}

The LAEs and random catalogs are shown in Figure~{\ref{fig:cat}}. To ensure the accuracy of the analysis, the random sources should match the areal selection of the data. We generated the random catalog used in this analysis from the ODIN narrowband-detected catalogs for each redshift. We first starmask the catalog to exclude sections that might be compromised by bright stars. We then make a data quality cut to the full catalog to include only data with existing HSC Deep broadband coverage. This creates a clover-shaped distribution of objects that matches that of the LAE catalog. Finally, we make a magnitude cut to only include sources within the central 90th percentile range of the narrowband magnitudes of the LAEs in our sample. The random catalog size varies for each redshift and is between 200--600 times the size of each LAE catalog. 

\subsection{Matter Correlation Function}

The clustering strength of galaxies can be compared to that of the underlying dark matter to learn about the properties of halos that host them. The matter angular auto-correlation function is a measure of the clustering of dark matter. In this analysis, we used the Core Cosmology Library \citep[\texttt{PyCCL}:][]{Chisari_2019} to determine the matter auto-correlation function corresponding to halos at the characteristic redshifts. The angular matter correlation function is dependent on the angular matter power spectrum, as given by the Limber equation 

\[
w_m(\theta) \approx \frac{1}{\pi} \int_0^\infty \frac{dz}{H(z)} \left( \frac{dN}{dz} \right)^2 P_m(\ell,\chi(z)),
\]
which simplifies the conversion between the real-space and redshift-space correlation functions \citep{Limber}. Here, $H(z)$ is the Hubble parameter at redshift $z$, $\frac{dN}{dz}$ is the expected redshift distribution of galaxies, $P_m$ is the matter power spectrum in Fourier space, $\ell$ is the angular multipole associated with the angular separation $\theta$, and $\chi(z)$ is the comoving distance at redshift $z$. We assume a nonlinear matter power spectrum to predict the galaxy auto-correlation function. The expected redshift distributions used for a$z=$2.4 and  $z=$3.1 are based on the clustering analysis conducted by \cite{White}, which utilized ODIN narrowband photometry at $z=$2.4 and $z=$3.1 with spectroscopic follow-up from the Dark Energy Spectrosocopic Instrument \citep[DESI;][]{DESI}. {We applied a correction to the \cite{White} expected redshift distributions to account for potential bias caused by the more conservative narrowband magnitude cut. To do this, we rescaled the expected redshift distribution by a factor of

\begin{equation}
{\frac{\text{FWHM of N(z) with ODIN mag. cut}}{\text{FWHM of N(z) with DESI mag. cut}}}
\end{equation} 

to obtain a slight increase in width expected for our dimmer cut. \cite{White} is discussed in more detail in Secion \ref{sec5.1}. We calculated the expected redshift distribution of LAEs at $z=$4.5 by first assuming the Schechter Luminosity function 

\begin{equation}\label{schecter}
    n(L) d(\log(L)) = \ln(10)\Phi^* \left(\frac{L}{L^*}\right)^{\alpha+1}\exp \left({\frac{-L}{L^*}}\right)d(\log(L))
\end{equation}
\citep{Schechter}. We assume $\text{log}\;L_*= 43.23$, $\text{log}\;\phi_* = -3.62$, and $\alpha = -2.00$ (G. Nagaraj et al., in prep). 
We converted the $5\sigma$ narrowband magnitude limit for the N673 filter (25.9) into a line flux limit at each wavelength using the filter transmission. We then determined the line luminosity limit from the line flux limit. Lastly, we then then calculated the expected redshift distribution, N($z$), by integrating the luminosity function above the line luminosity limit. The expected redshift distributions used in the analysis are shown in Figure \ref{fig:Nz}. 

\begin{figure*}
\includegraphics[width=\textwidth]{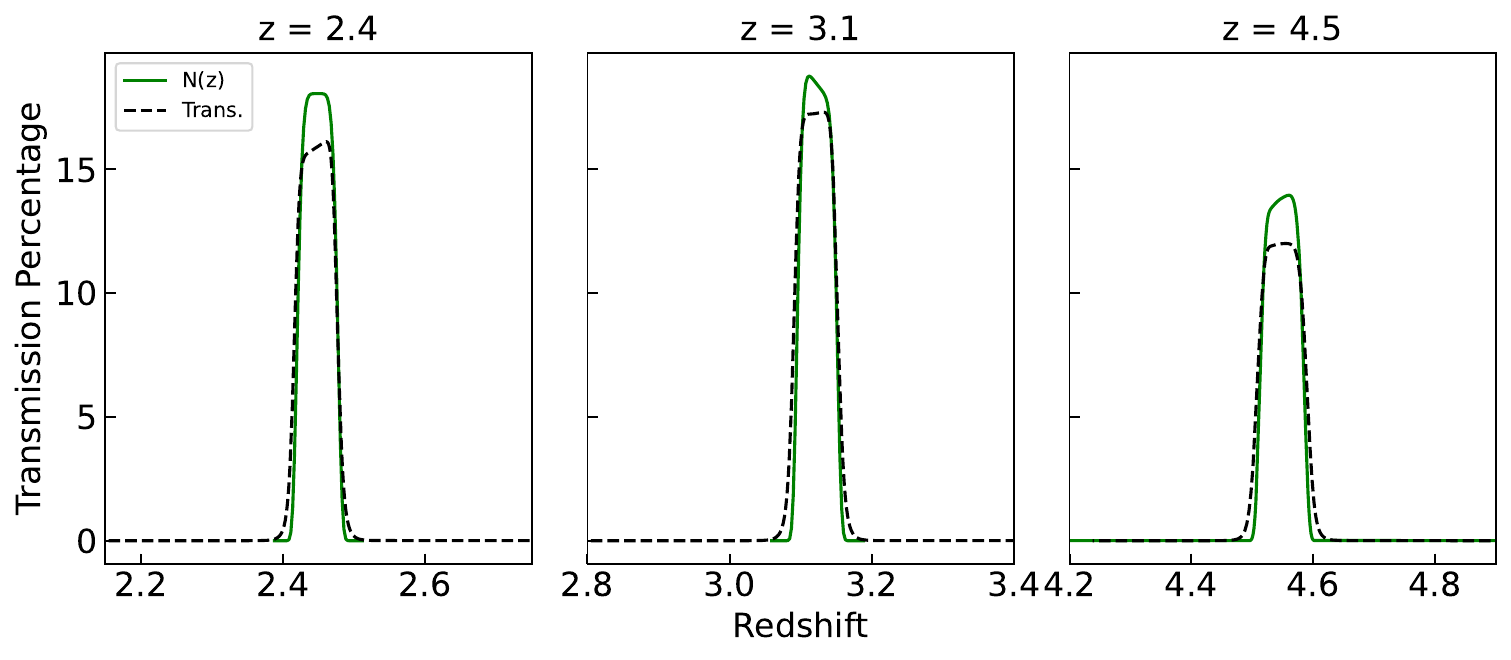}
\caption{The expected redshift distribution, N(z), of LAEs plotted over the transmission curve for the three redshift samples. The relative strength of the transmission curve is analogous to the percentage of total transmission. For N(z), this refers to the relative probability of detecting LAEs at some redshift. For $z=$2.4 and $z=$3.1 we use a \textbf{corrected} DESI-spectroscopy based N(z) from \cite{White}, while for $z = $4.5 we use a prediction based on the filter transmission curve and the LAE luminosity function.} 
\label{fig:Nz}
\end{figure*}

\begin{figure}[]
    \hspace*{-1cm}
   \includegraphics[width=0.5\textwidth]{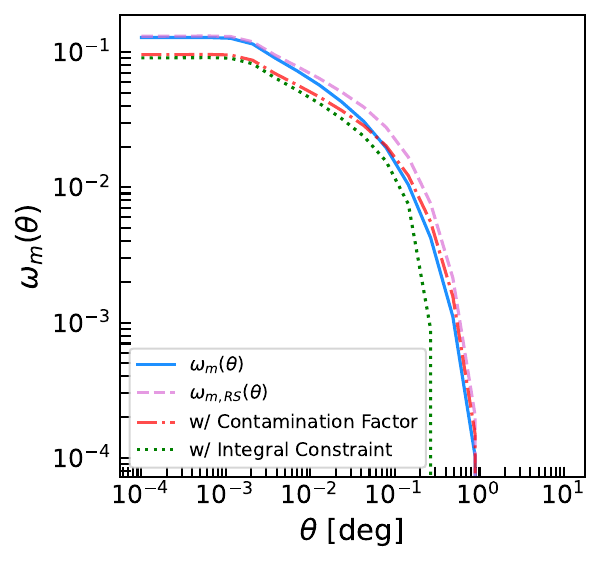}
    \caption{The series of systematic corrections applied to $w_m(\theta)$ at $z = $3.1 for galaxy bias of 1. The corrections are applied sequentially such that the dotted green line includes all corrections. The solid blue line is the $w_m(\theta)$ obtained from \texttt{PyCCL}. The dashed pink line corrects this model for redshift-space distortions. The dash-dotted red line adds a multiplicative contamination correction to the dashed pink line. Finally, the dotted green line includes all of the previous corrections and accounts for the integral constraint.}This line is the corrected $\omega_{model}(\theta)$ that is used for the analysis.
\label{fig:corr}
\end{figure}

\subsection{Modeling the Observed Correlation Function}

The matter angular correlation function, $\omega_m(\theta)$, was produced in real space using  \texttt{PyCCL}, but it does not account for any observational effects. Therefore, we applied a series of corrections to the \texttt{PyCCL} calculation of the real-space matter correlation function, as shown in Figure~\ref{fig:corr}. We note that there is a systematic uncertainty on $w_m(\theta)$ from the width of the expected redshift distribution and from the cosmological parameters, but we do not address this because the uncertainties are negligible when compared to the amplitude of the jacknife uncertainties on $w_{LS}(\theta)$.

First, we corrected for redshift-space distortions following the procedure presented in \cite{White}. Galaxy peculiar velocities cause redshift-space distortions (RSD), which shift the apparent radial position of the galaxy \citep{Hamilton}. On large scales, the infall of galaxies toward overdensities causes their distribution to appear compressed. This affects the clustering because it shifts the features from a broader redshift range into our observed redshift range. We address these distortions by applying corrections to the matter angular correlation function.  
\citet{White} first produced the RSD-corrected correlation function from simulation-based mock catalogs. They then integrated the catalogs through the LAE selection function to predict the measured real-space clustering. Lastly, they used a projection integral to turn the resulting 3D clustering signal into the observed angular clustering, integrated across the redshift distribution of ODIN. 
We used this model to apply a multiplicative correction to the \texttt{PyCCL} $\omega_m(\theta)$ to obtain $\omega_{m,RS}(\theta)$. 

Next, we applied a contamination correction based on ongoing spectroscopic follow-up of ODIN LAEs (E. Pinarski et al., in prep). We compare spectroscopic to photometric data to define a contamination fraction: 
\begin{equation}
    f = \left(\frac{\textnormal{Objects with spectra \textbf{NOT} identified as LAEs}}{\textnormal{Total objects with spectra}}\right)
    \label{eq10}
\end{equation}
 Typical contaminants include broad and narrow-line AGN, foreground galaxies, and sources with undetermined redshift.  These undetermined sources could be spurious detections, continuum-only galaxies, or real LAEs whose emission lines have signal-to-noise insufficient for a spectroscopic detection in that particular fiber.  Since the most common contaminants to LAE samples with observed equivalent widths greater than 80 {\AA} are continuum-only objects exhibiting spurious narrowband excess \citep{Gawiser07}, it is reasonable to assume that the contaminants are only weakly clustered. This reduces the strength of $\omega_g(\theta)$ by a factor of $(1-f)^2$.  Our initial spectroscopy (E. Pinarski et al., in prep) finds contamination factors of 0.22, 0.11, and 0.14 for $z=$2.4, 3.1, and 4.5 LAEs, respectively. 
Because the spectroscopy does not yet go quite as deep as our LAE catalogs and given the uncertaint interpretation of undetermined sources, 
  we assume an uncertainty of $\pm$0.1 on the contamination fraction at each redshift.

Galaxy bias, $b$, describes how the density fluctuations of luminous matter compare to the density of the dark matter in the universe. It is represented as a multiplicative factor between the galaxy and dark matter angular correlation function. Linear galaxy bias is defined as 

\begin{equation}
    b = \left(\frac{\omega_g({\theta})}{\omega_m({\theta})}\right)^{\frac12}
\end{equation}
where $\omega_g(\theta)$ is the LAE angular correlation function and $\omega_m(\theta)$ is the matter angular correlation function. 

The final correction was the inclusion of the integral constraint, such that 

\begin{equation}
    \omega_{\text{model}}(\theta) = \frac{b^2(1-f)^2\omega_{m,RS}(\theta) - \omega_{\Omega}(b,f)}{1 + \omega_{\Omega}(b,f)}
    \label{model}
\end{equation}
where the integral constraint is defined as

\begin{equation}
     \omega_{\Omega}(b,f) = \frac{\sum RR(\theta) b^2(1-f)^2 \omega_{m,RS}(\theta)}{\sum RR(\theta)}
     \label{eq9}
\end{equation}
$b$ refers to galaxy bias, and $f$ is the contamination fraction given above. The final angular correlation function with all of the applied corrections will be referred to as $\omega_{\text{model}}(\theta)$. 
In order to exclude small scales dominated by galaxy pairs within a single halo, which are difficult to model, the fitting range for the analysis starts at the one-to-two halo transition at separations of about $0.004^\circ$.  The fitting range ends at $1 ^\circ$, above which the redshift-space distortion correction starts to become too large to be reliable.

\qquad 
We use Equation \ref{model} to calculate the best-fit value of bias. The uncertainties were calculated by assuming a uniform prior on bias and performing a Bayesian analysis to find the bias values within a 68\% credible region. We take this credible region as the uncertainty in the product $b(1-f)$, since the product of linear bias factor and purity is what appears in the model, and we add the impact from the 10\% uncertainty in contamination fraction noted above in quadrature to generate the final uncertainty in bias. We are assuming that there are formally no uncertainties on $w_m(\theta)$.

\subsection{Dark Matter Halo Mass}
Once we have the galaxy bias factor, we can use \texttt{CCL} to calculate the typical mass of a dark matter halo hosting an LAE. We use the matter power spectrum at the redshift of the LAE sample of interest and the galaxy bias function, which describes galaxy bias as a function of halo mass and redshift \citep{Tinker2010}. Assuming that all dark matter halos above some minimum mass host LAEs, we find the minimum halo mass that corresponds to the observed galaxy bias.  

 We used a number-weighted bias 
\begin{equation}\label{numberweightedbias}
    b=\frac{\int_{M_{\text{min}}}^{\infty}b(M)n(M)dM}{\int_{M_{\text{min}}}^{\infty}n(M)dM}
\end{equation}
 where $b(M)$ is the galaxy bias function and $n(M)$ is the galaxy mass function, to vary the minimum halo mass, $M_{min}$, until the number weighted halo bias matched the observed best-fit LAE bias. 
 We also determined the median halo mass of this set of halos.  
 Although there could be many ways to assign LAEs to halos and recover the observed bias value, the median halo mass is quite similar across this range of halo occupation distributions, making it a robust quantity.  
We calculated uncertainties by finding the minimum and median halo masses that corresponded to the upper and lower 68\% confidence bounds of galaxy bias. 

\begin{figure*}[!htb]
   \includegraphics[width=18cm]{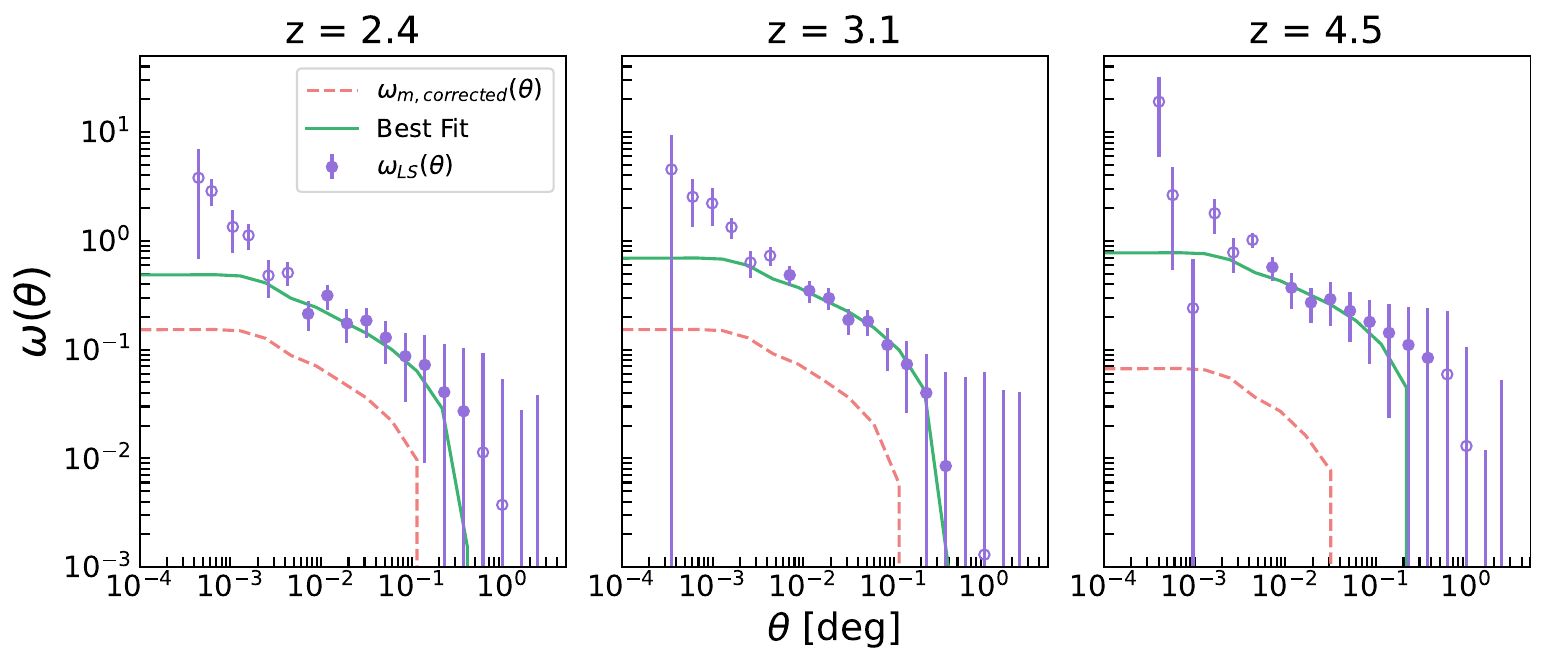}
    \caption{A comparison of the observed LAE angular correlation function ($\omega_{LS}(\theta)$), the corrected matter angular correlation function ($\omega_{m,corrected}(\theta)$), and the best-fit model for the LAE angular correlation function. The filled points represent our fitting range $(0.004^\circ < \theta < 1^\circ)$, which was used to calculate bias and 
    corresponding halo properties.  Open circles show datapoints outside this range that were therefore excluded from the fit. The smallest angles correspond to the one-halo regime, which requires additional modeling to conduct an accurate analysis. The one-halo contribution is clearly detected, as the current model fails to account for the small-scale excess.
    Datapoints at $\theta > 1 ^\circ$ are excluded because in this range the redshift space distortions have a larger effect, making the correction less certain.} 
    \label{fig:w}
\end{figure*}

\subsection{Halo Occupation Fraction}

The halo occupation fraction (HOF) is defined as the ratio of the LAE number density to the number density of dark matter halos with masses greater than the minimum mass required to host an LAE. 
The latter is calculated in the integral in the denominator in equation \ref{numberweightedbias}. The former is calculated by dividing the number of LAEs in each redshift catalog by the volume of the ODIN COSMOS field. 
However, incompleteness in the survey means that  the number of LAEs at the faint end is underestimated. To address this, we performed a number density correction. We fit a power law to binned LAE narrowband magnitudes and calculated the completeness of the sample. We selected a cutoff magnitude where the power law fit overestimates the number density, indicating incompleteness in the LAE catalog. We selected cutoff magnitudes corresponding to 75-80\% completeness. 
The correction reduced the LAE sample sizes from 6100, 5782 and 4101 to 
final samples of 5233, 5220, and 3706, respectively, at $z=$2.4, 3.1, and 4.5. 
These final samples were used for our clustering analysis.  
We estimated uncertainties in the HOFs by propagating the uncertainties in halo number density while noting that Poisson uncertainties in LAE number density  are negligible in comparison. 

\section{Results}\label{sec4}

\subsection{Clustering Properties of LAEs}

\begin{table*}[ht]
\centering
\caption{Summary of Parameters for LAE and Halo Properties}
\label{tab:lae_halo_properties}
\footnotesize  
\renewcommand{\arraystretch}{2.4}
\hspace*{-4.0cm}
\begin{threeparttable}
\begin{tabular}{@{\hskip 3pt}l@{\hskip 3pt}c@{\hskip 3pt}c@{\hskip 3pt}c@{\hskip 3pt}c@{\hskip 3pt}c@{\hskip 3pt}c@{\hskip 3pt}c@{\hskip 3pt}c@{\hskip 3pt}c@{\hskip 3pt}c@{\hskip 3pt}}
\toprule
\textbf{\(z\)} 
 & \textbf{Cut\tnote{*}} 
 & \textbf{\(  L_{Ly\alpha,\lim}\)\(^\diamond\)}
 & \textbf{\(\rm n_{LAE}\) [cMpc\(^{-3}\)]} 
 & \textbf{Compl.\(^\dagger\)} 
 & \textbf{Purity} 
 & \textbf{Bias} 
 & \textbf{\(\log(M_{h,min}/M_{\odot})\)} 
 & \textbf{\(\log(M_{h,med}/M_{\odot})\)} 
 & \textbf{\(\rm n_{halo}\)[cMpc\(^{-3}\)]} 
 & \textbf{HOF} \\
\midrule
2.4 & 25.2 & \(2.04 \times 10^{42}\) & \(5.99 \pm 0.07 \times 10^{-4}\) & 0.79 & \(0.77 \pm 0.10\) & \(1.72^{+0.26}_{-0.27}\) & \(11.19^{+0.32}_{-0.30}\) & \(11.44^{+0.30}_{-0.28}\) & \(0.8^{+0.9}_{-0.5} \times 10^{-2}\) & \(0.067^{+0.078}_{-0.039}\) \\
3.1 & 25.3 & \(2.33 \times 10^{42}\) & \(6.51 \pm 0.07 \times 10^{-4}\) & 0.77 & \(0.88 \pm 0.10\) & \(2.01^{+0.26}_{-0.29}\) & \(10.89^{+0.27}_{-0.27}\) & \(11.13^{+0.26}_{-0.26}\) & \(1.3^{+1.4}_{-0.7} \times 10^{-2}\) & \(0.047^{+0.051}_{-0.025}\) \\
4.5 & 25.5 & \(3.49 \times 10^{42}\) & \(3.90 \pm 0.05 \times 10^{-4}\) & 0.76 & \(0.85 \pm 0.10\) & \(2.95^{+0.40}_{-0.46}\) & \(10.64^{+0.25}_{-0.25}\) & \(10.85^{+0.24}_{-0.24}\) & \(1.1^{+1.3}_{-0.6} \times 10^{-2}\) & \(0.032^{+0.037}_{-0.018}\) \\
\bottomrule
\end{tabular}
    \begin{tablenotes}
        \raggedleft
        \small{
        \item {$\diamond$ Ly$\alpha$ Luminosity Limit [erg,s$^{-1}$]}
        \item {$^*$ Narrow-band AB magnitude cutoff} 
        \item {$\dagger$ Completeness}
        }
    \end{tablenotes}
\end{threeparttable}
\end{table*}

\begin{figure*}[htb!]
\centering
\includegraphics[width=18cm]{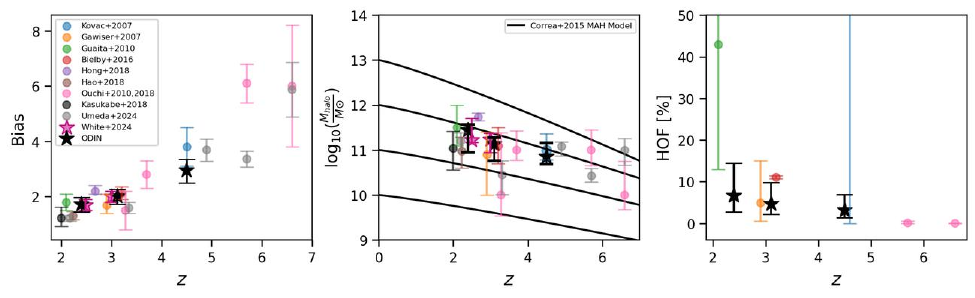}
\caption{The redshift evolution of three LAE halo properties from our analysis compared to results from the literature. The colored circles represent results from the literature \citep{Kova__2007,Gawiser07,Guaita10,Bielby,Hong_2019,Hao,Ouchi10,Ouchi18,Kasukabe,umeda} and the stars represent results that utilize ODIN data. The open stars represents results from the \cite{White} DESI spectroscopy-based LAE clustering analysis, while the filled stars represent the results from this analysis. The left panel shows the evolution of galaxy bias values reported by different studies. The redshift values of \citealt{Gawiser07}, \citealt{Ouchi10}, \citealt{Bielby}, \citealt{umeda}, and \citealt{White} were shifted around $z =$ 3.1 to make the bias points more visible. The middle panel shows the evolution of halo mass, and compares it to the halo mass assembly history predicted by the \cite{Correa} model. The halo mass values of \citep{Hong_2019,Hao,Kasukabe,umeda,White} were estimated from the bias measurements using our methodology. The right panel shows the halo occupation fraction (HOF) of LAEs, with \cite{Kova__2007} shifted by 0.1 to make it visible. HOF values are only shown for those studies that reported the completeness values that provide a needed input for this quantity.}
\label{fig:halo_evol}
\end{figure*}

We conducted a clustering analysis of over 14,000 ODIN LAEs at $z = $2.4, 3.1, and 4.5 and determined properties of the LAEs and their corresponding halos. The resulting correlation functions are shown in Figure~\ref{fig:w}, with halo properties summarized in Table \ref{tab:lae_halo_properties}. The analysis revealed galaxy bias values that increased with redshift, which is consistent with expectations from hierarchical structure formation because halos of similar mass at are less common at higher redshifts. We significantly reduced uncertainties compared to previous studies numbers of LAEs at each redshift, as shown in the left panel of Figure \ref{fig:halo_evol}.  

Using the best fit bias values, we calculated minimum and median halo mass required to host an LAE. The minimum halo masses were between $10^{10.6}-10^{11.2} M_{\odot}$. We calculated the median halo masses above this minimum and found values of $10^{10.8}-10^{11.4} M_{\odot}$. We used this method to estimate the median halo masses for studies that did not explicitly report them but did report bias values, as shown in Figure \ref{fig:halo_evol}. We did not correct other surveys’ results for modest differences between our assumed value of $\sigma_8$. We integrated above the minimum halo mass to calculate the number density of halos with masses greater than the minimum required to reproduce the LAE bias and found halo number densities between 0.8--1.29 $\times $ $10^{-2}$ cMpc$^{-3}$. Combining this with the LAE number densities, we found halo occupation fractions of roughly 3--7\%. One important caveat is that these halo occupation fractions result from the very simple assumption that each halo above the minimum mass hosts exactly one LAE. Pending a detailed halo occupation distribution, this creates significant systematic uncertainty in the halo occupation fractions. However, the results for galaxy bias and median halo mass are determined by the observed clustering strength and therefore are robust to uncertainties in the halo occupation fraction.

\subsection{Evolution of LAE Clustering Bias, Halo Masses and Halo Occupations}

We use our observations of various redshift slices to study the evolution of the properties of LAEs and their host halos. The bias of a galaxy population is expected to decrease over time because dark matter becomes more clustered as dense regions are gravitationally pulled together and halos grow in mass. This decreases the relative clustering of the LAEs. Our bias results are consistent with this expectation, as shown in Figure~\ref{fig:halo_evol}. 

The results indicated slight evolution in minimum and median halo mass across the three redshifts. Host halo mass seems to decrease as redshift increases, indicating that LAEs reside in larger halos in the present. This is also consistent with the expectation that dark matter becomes more clustered over cosmic time. The evolution of median halo mass aligns reasonably well with the model curve for a present-day  $10^{11}M_{\odot}$ halo. 

The halo occupation fraction of the LAEs increases from 3\% to 7\% across the three redshift samples. Given the  uncertainties propagated from halo number density, the ODIN results are consistent with no evolution.  However, once the $z>5$ results from \citet{Ouchi18} and $z=2.1$ result from \citet{Guaita10} are included,  Fig.~\ref{fig:halo_evol} shows evidence for a decreasing HOF with redshift.    


\subsection{Connection to Present Day Galaxies}


The evolution of halo properties revealed by our analysis is consistent with current models of structure formation. The right panel of Figure \ref{fig:halo_evol} shows the evolution of halo mass required to host an LAE as reported in this study and others in the literature \citep[][]{Kova__2007, Gawiser07, Guaita10, Bielby, Hong_2019, Hao, Ouchi10,Ouchi18, Kasukabe, umeda}. 
The values are consistent with the \cite{Correa} halo mass assembly model curves for a present-day halo just under $10^{12} M_{\odot}$. The results suggest that LAEs are progenitors of modern galaxies with halo masses between $10^{11} M_{\odot} \textnormal{ and } 10^{12} M_{\odot}$, corresponding to Large Magellanic Cloud (LMC)-like to Milky Way-like galaxies \citep{LMC}. 

 \section{Discussion}\label{sec5}

\subsection{Comparison to DESI-ODIN results}\label{sec5.1}

\cite{White} performed a similar analysis using  samples of 2382 LAEs at $z = $ 2.4 and 1956 LAEs at $z = $ 3.1 in the COSMOS field. The LAE samples were selected by first creating a ``liberal" photometric catalog that allowed for a higher fraction of interlopers, initially prioritizing completeness over sample purity and then performing spectroscopy. The final LAE samples were created by  applying narrowband magnitude and color cuts to ensure high purity of the samples and a well-measured contamination rate. These samples overlap significantly with the ODIN LAE samples analyzed here because these were selected from the same deep narrowband images. However, the sets of galaxies used in this paper are more than twice as large as the more conservative final sample selected by \citeauthor{White}. Nonetheless, the results from the two analyses are in remarkable agreement. 
\cite{White} reported LAE number densities around $10^{-3} h^3 \text{cMpc}^{-3}$ for both of their redshift samples. They calculated galaxy bias values of 1.7 $\pm$ 0.2 and 2.0 $\pm$ 0.2 respectively for $z = $2.4 and $z=$3.1, which are a close match to the bias values shown in Table~\ref{tab:lae_halo_properties}.   

\subsection{
Why is the LAE Halo Occupation Fraction So Low?}

The LAE halo occupation fractions were roughly 3--7\%. This is consistent with existing measurements of LAE halo occupation fractions in our redshift range \citep{Gawiser07,Bielby,Kova__2007}, although such measurements were sparse and often affected by significant uncertainties. These results indicate that few halos above the minimum mass of $\sim 10^{10.8}$M$_\odot$ undergo star formation episodes that produce detectable Ly$\alpha$ emission at $2.4<z<4.5$. 
Potential explanations for this low halo occupation fraction should consider the escape of Ly$\alpha$ photons from the galaxy and nearby IGM as well as the ``duty cycle'' i.e., the fraction of time spent in a low-dust starburst.
For example, the low HOF might imply that Ly$\alpha$ emission is a temporary phase of galaxy evolution characterized by high star formation rates, and only a fraction of galaxies are undergoing this phase at any given time. It is also possible that LAEs do not emit isotropically. In the latter scenario, the Ly$\alpha$ photons efficiently escape each galaxy only in preferred directions. The observed HOF then implies that the typical opening angle for Ly$\alpha$ escape might be as small as $(4 \pi)/30$ steradians. \cite{Momose} found that LAEs have anisotropic cross correlations with the Ly$\alpha$ forest transmission fluctuations such that the average HI density is higher on the more distant side of the galaxy. 
This implies that LAEs are less likely to be detected when a denser IGM environment is present on the near side along our line of sight. 
The three factors described above (the fraction of halos hosting galaxies experiencing a low-dust starburst, the fraction of such galaxies emitting a high Ly$\alpha$ luminosity towards Earth, and the fraction of the latter with a sufficiently low-density IGM in front to transmit those photons) should multiply to produce the observed HOF.

Low LAE HOFs are more commonly reported at higher redshifts  \citep[e.g.,][]{ Ouchi18}, likely due to decreasing Ly$\alpha$ visibility or shorter HOFs. The similarly low values we find at $z = 2.4-4.5$, though less expected, may suggest that the same processes are limiting the detectability of LAEs at these redshifts. One way to determine if the low HOF is surprising is to start with the assumption that LAEs are a subset of UV-selected (Lyman break) galaxies.  If so, the LAE HOF should be a product of the UV-selected galaxy HOF and the fraction of UV-selected galaxies exhibiting sufficient Ly$\alpha$ equivalent width to be classified as LAEs. The HOFs of UV-selected galaxies have been measured in several studies, including \citet{Lee2009}, who looked at LBG HOFs for one of the dimmest samples yet studied and found 
that LBGs typically occupy less massive halos than our LAE samples.
For the latter factor, the equivalent width threshold is important because lowering it would allow galaxies with weak  Ly$\alpha$ emission to be classified spectroscopically as LAEs, inflating the observed LAE fraction of LBGs. This would lead to an overestimation of the LAE HOF \citep{Caruana18}. 
\cite{Shapley} and \cite{Stark} used UV-selected ($m_{UV} \leq 25$-26) galaxy samples
 at $z\sim3$ with a comparable rest-frame equivalent width threshold (20-25$\AA$) and found that a  
 significant fraction  ($25-50$\%) of continuum-selected star-forming galaxies at these redshifts would be classified as LAEs.
For a sample of LBGs at $z = 3-6$ with $m_{UV} \simeq 27$, comparable to the redshift and UV magnitude range of our LAEs, \cite{Kusakabe2020} found an LAE fraction of $20-40$\% while requiring a Ly$\alpha$  equivalent width threshold  of 25$\AA$. 
   
When multiplied with the high HOF ($\sim60^{+20}_{-30}$\%) found for LBGs at $z \sim 4$ with $m_{UV} = 25-30 $ \citep{Harikane2016}, this predicts an LAE HOF in the range of 6-32\%, 
which is only consistent with our findings for the LAE samples at the low end of the range. 
Notably, if the lower bound of the LBG HOF uncertainty is taken into account, the resulting LAE HOF estimate shifts closer to 10\%, which is statistically consistent with our measured $z$ = 3.1 and 4.5 within 1 to 2$\sigma$.

Even if future measurements reduce the uncertainties to generate a clear prediction of an LAE HOF $>$15\%, there would be a possible resolution to the discrepancy. 
Specifically, our low HOFs could be caused by incompleteness in narrowband-selected LAE samples. Narrowband detection is subject to observational constraints including equivalent width cuts, continuum detection thresholds, and filter transmission limits. 
This incompleteness can be determined via comparison with blind spectroscopic surveys such as HETDEX. This would explain the relatively low HOFs found in this study as well as previous results from photometric surveys \citep{Kova__2007,Bielby,Gawiser07}. 
}

\subsection{How long-lived is the LAE phenomenon?}
If Ly$\alpha$ emission is indeed a temporary phase of galaxy evolution, we can use the LAE halo occupation fraction to approximate the duration of the LAE phase. If we make the simplifying assumption that each galaxy undergoes exactly one Ly$\alpha$ emission phase during its lifetime, we can estimate the amount of time this phase would need to last to produce our halo occupation fraction. The LAE phase would need to last a fraction of the Hubble time i.e., roughly $\text{HOF}(z)\times t_H(z)$, which is 40-180 million years at our three redshifts. This is consistent with LAE timescales in the literature. \cite{Gawiser07} found that $z \approx 3.1$ LAEs have stellar ages of around 100 million years. Similarly, \cite{Cowie} estimated that the LAE lifetimes of $z<0.4$ LAEs are about 110 million years. \cite{Nicole25} determined the mass-weighted median stellar ages, $t_{50}$, of LAEs and found median values of 80, 250, and 240 Myr for $z$ = 2.4, 3.1, and 4.5. These results indicate that our estimated LAE timescales are a plausible match for observed stellar ages, so there is no obvious inconsistency with this highly simplified model. 
%
However, this motivates further investigation of whether galaxies can remain in an LAE phase for that time period without acquiring enough dust to attenuate their Ly$\alpha$ emission significantly. 

\section{Conclusion}\label{conc}
We conducted a clustering analysis of approximately 14,000 LAEs at $z = $ 2.4, 3.1, and 4.5. We used a catalog of LAEs selected from the ODIN survey in the COSMOS field to investigate the properties of LAEs and their host halos. To perform the analysis, we used \texttt{TreeCorr} to determine the LAE angular correlation function, $\omega_g(\theta)$. Then, we used \texttt{PyCCL} and the LAE expected redshift distribution, N($z$), to model the matter angular correlation function, $\omega_m(\theta)$. Lastly, we corrected $\omega_m(\theta)$ to account for various cosmic effects. 

The analysis revealed galaxy bias values of $1.71^{+0.26}_{-0.27},  2.01^{+0.26}_{-0.29}, \textnormal{ and } 2.95^{+0.40}_{-0.46}$ for $z$ = 2.4, 3.1, and 4.5, respectively, corresponding to median halo mass values of $11.44^{+0.30}_{-0.28},  11.13^{+0.26}_{-0.26}, \textnormal{ and } 10.85 ^{+0.24}_{-0.24}\textnormal{ log}(M_{\odot})$, and halo occupation fractions of $0.067^{+0.078}_{-0.039}, 0.047^{+0.051}_{-0.025}, \textnormal{ and}$

$0.032^{+0.037}_{-0.018}$, respectively. 
The evolution of halo mass is consistent with the idea that LAEs are progenitors of typical present-day galaxies, since their descendants are predicted to reside in halos with masses between LMC-like and Milky-Way-like galaxies. The relatively low halo occupation fraction results suggest that 
strong Ly$\alpha$ emission is a temporary, low-dust phase of galaxy evolution. 

In-depth studies of the star formation histories of LAEs, as well as the correlation between these histories and their dark matter halo masses, should help to assess the duration and ubiquity of the LAE phenomenon. Future ODIN analyses will contain a sample of over 100,000 LAEs across 7 cosmic fields, allowing for tighter constraints on LAE and halo properties and enabling further investigation of the topics explored in this paper. 

\section{Acknowledgments}

This work utilizes observations at Cerro Tololo Inter-
American Observatory, NSFs NOIRLab (Prop. ID
2020B-0201; PI: KSL), which is managed by the Association of Universities for Research in Astronomy under
a cooperative agreement with the NSF.  DH, EG, and NF 
acknowledge support from NSF grant AST-2206222, and DH would like to acknowledge related support from the NSF Alliances for Graduate Education and the Professoriate Graduate Research Supplements (AGEP-GRS).  
EG also acknowledges the support of an
IBM Einstein fellowship for his sabbatical at IAS during the completion of this manuscript. KSL and VR acknowledge financial support from the NSF under Grant. Nos. AST-2206705 and AST-2408359 and from the Ross-Lynn Purdue Research Foundations.
YY is supported by the Basic Science Research Program through the National Research Foundation of Korea funded by the Ministry of Science, ICT \& Future
Planning (2019R1A2C4069803). NF acknowledges the support of NSF Graduate Research Fellowship Program under Grant No. DGE-2233066. RC and CG acknowledge support from NSF grant AST-2408358. The Institute for Gravitation and the Cosmos is supported by
the Eberly College of Science and the Office of the Senior Vice President for Research at the Pennsylvania
State University. LG thanks support from FONDECYT regular proyecto No. 1230591. HSH acknowledges the support of
the National Research Foundation of Korea grant, No.
2022R1A4A3031306, funded by the Korean government
(MSIT).


\newpage

\bibliography{main}{}

\begin{thebibliography}{}
\expandafter\ifx\csname natexlab\endcsname\relax\def\natexlab#1{#1}\fi
\providecommand{\url}[1]{\href{#1}{#1}}
\providecommand{\dodoi}[1]{doi:~\href{http://doi.org/#1}{\nolinkurl{#1}}}
\providecommand{\doeprint}[1]{\href{http://ascl.net/#1}{\nolinkurl{http://ascl.net/#1}}}
\providecommand{\doarXiv}[1]{\href{https://arxiv.org/abs/#1}{\nolinkurl{https://arxiv.org/abs/#1}}}

\bibitem[{{Aihara} {et~al.}(2018){Aihara}, {Arimoto}, {Armstrong}, {Arnouts}, {Bahcall}, {Bickerton}, {Bosch}, {Bundy}, {Capak}, {Chan}, \& et~al.}]{SubaruHSC}
{Aihara}, H., {Arimoto}, N., {Armstrong}, R., {et~al.} 2018, \pasj, 70, S4, \dodoi{10.1093/pasj/psx066}

\bibitem[{{Bertin} \& {Arnouts}(1996)}]{SE}
{Bertin}, E., \& {Arnouts}, S. 1996, \aaps, 117, 393, \dodoi{10.1051/aas:1996164}

\bibitem[{{Bielby} {et~al.}(2016){Bielby}, {Tummuangpak}, {Shanks}, {Francke}, {Crighton}, {Ba{\~n}ados}, {Gonz{\'a}lez-L{\'o}pez}, {Infante}, \& {Orsi}}]{Bielby}
{Bielby}, R.~M., {Tummuangpak}, P., {Shanks}, T., {et~al.} 2016, \mnras, 456, 4061, \dodoi{10.1093/mnras/stv2914}

\bibitem[{{Caruana} {et~al.}(2018){Caruana}, {Wisotzki}, {Herenz}, {Kerutt}, {Urrutia}, {Schmidt}, {Bouwens}, {Brinchmann}, {Cantalupo}, {Carollo}, {Diener}, {Drake}, {Garel}, {Marino}, {Richard}, {Saust}, {Schaye}, \& {Verhamme}}]{Caruana18}
{Caruana}, J., {Wisotzki}, L., {Herenz}, E.~C., {et~al.} 2018, \mnras, 473, 30, \dodoi{10.1093/mnras/stx2307}

\bibitem[{Chisari {et~al.}(2019)Chisari, Alonso, Krause, Leonard, Bull, Neveu, Villarreal, Singh, McClintock, Ellison, Du, Zuntz, Mead, Joudaki, Lorenz, Tröster, Sanchez, Lanusse, Ishak, Hlozek, Blazek, Campagne, Almoubayyed, Eifler, Kirby, Kirkby, Plaszczynski, Slosar, Vrastil, \& and}]{Chisari_2019}
Chisari, E., Alonso, D., Krause, E., {et~al.} 2019, The Astrophysical Journal Supplement Series, 242, 2, \dodoi{10.3847/1538-4365/ab1658}

\bibitem[{{Correa} {et~al.}(2015){Correa}, {Wyithe}, {Schaye}, \& {Duffy}}]{Correa}
{Correa}, C.~A., {Wyithe}, J. S.~B., {Schaye}, J., \& {Duffy}, A.~R. 2015, \mnras, 450, 1514, \dodoi{10.1093/mnras/stv689}

\bibitem[{{Cowie} {et~al.}(2011){Cowie}, {Barger}, \& {Hu}}]{Cowie}
{Cowie}, L.~L., {Barger}, A.~J., \& {Hu}, E.~M. 2011, \apj, 738, 136, \dodoi{10.1088/0004-637X/738/2/136}

\bibitem[{{DESI Collaboration} {et~al.}(2016){DESI Collaboration}, {Aghamousa}, {Aguilar}, {Ahlen}, {Alam}, {Allen}, {Allende Prieto}, {Annis}, {Bailey}, {Balland}, {Ballester}, {Baltay}, {Beaufore}, {Bebek}, {Beers}, {Bell}, {Bernal}, {Besuner}, {Beutler}, {Blake}, {Bleuler}, {Blomqvist}, {Blum}, {Bolton}, {Briceno}, {Brooks}, {Brownstein}, {Buckley-Geer}, {Burden}, {Burtin}, {Busca}, {Cahn}, {Cai}, {Cardiel-Sas}, {Carlberg}, {Carton}, {Casas}, {Castander}, {Cervantes-Cota}, {Claybaugh}, {Close}, {Coker}, {Cole}, {Comparat}, {Cooper}, {Cousinou}, {Crocce}, {Cuby}, {Cunningham}, {Davis}, {Dawson}, {de la Macorra}, {De Vicente}, {Delubac}, {Derwent}, {Dey}, {Dhungana}, {Ding}, {Doel}, {Duan}, {Ealet}, {Edelstein}, {Eftekharzadeh}, {Eisenstein}, {Elliott}, {Escoffier}, {Evatt}, {Fagrelius}, {Fan}, {Fanning}, {Farahi}, {Farihi}, {Favole}, {Feng}, {Fernandez}, {Findlay}, {Finkbeiner}, {Fitzpatrick}, {Flaugher}, {Flender}, {Font-Ribera}, {Forero-Romero}, {Fosalba}, {Frenk}, {Fumagalli}, {Gaensicke}, {Gallo},
  {Garcia-Bellido}, {Gaztanaga}, {Pietro Gentile Fusillo}, {Gerard}, {Gershkovich}, {Giannantonio}, {Gillet}, {Gonzalez-de-Rivera}, {Gonzalez-Perez}, {Gott}, {Graur}, {Gutierrez}, {Guy}, {Habib}, {Heetderks}, {Heetderks}, {Heitmann}, {Hellwing}, {Herrera}, {Ho}, {Holland}, {Honscheid}, {Huff}, {Hutchinson}, {Huterer}, {Hwang}, {Illa Laguna}, {Ishikawa}, {Jacobs}, {Jeffrey}, {Jelinsky}, {Jennings}, {Jiang}, {Jimenez}, {Johnson}, {Joyce}, {Jullo}, {Juneau}, {Kama}, {Karcher}, {Karkar}, {Kehoe}, {Kennamer}, {Kent}, {Kilbinger}, {Kim}, {Kirkby}, {Kisner}, {Kitanidis}, {Kneib}, {Koposov}, {Kovacs}, {Koyama}, {Kremin}, {Kron}, {Kronig}, {Kueter-Young}, {Lacey}, {Lafever}, {Lahav}, {Lambert}, {Lampton}, {Landriau}, {Lang}, {Lauer}, {Le Goff}, {Le Guillou}, {Le Van Suu}, {Lee}, {Lee}, {Leitner}, {Lesser}, {Levi}, {L'Huillier}, {Li}, {Liang}, {Lin}, {Linder}, {Loebman}, {Luki{\'c}}, {Ma}, {MacCrann}, {Magneville}, {Makarem}, {Manera}, {Manser}, {Marshall}, {Martini}, {Massey}, {Matheson}, {McCauley}, {McDonald},
  {McGreer}, {Meisner}, {Metcalfe}, {Miller}, {Miquel}, {Moustakas}, {Myers}, {Naik}, {Newman}, {Nichol}, {Nicola}, {Nicolati da Costa}, {Nie}, {Niz}, {Norberg}, {Nord}, {Norman}, {Nugent}, {O'Brien}, {Oh}, \& {Olsen}}]{DESI}
{DESI Collaboration}, {Aghamousa}, A., {Aguilar}, J., {et~al.} 2016, arXiv e-prints, arXiv:1611.00036, \dodoi{10.48550/arXiv.1611.00036}

\bibitem[{{Firestone} {et~al.}(2024){Firestone}, {Gawiser}, {Ramakrishnan}, {Lee}, {Valdes}, {Park}, {Yang}, {Ciardullo}, {Artale}, {Benda}, {Broussard}, {Eid}, {Farooq}, {Gronwall}, {Guaita}, {Gwyn}, {Hwang}, {Im}, {Jeong}, {Karthikeyan}, {Lang}, {Moon}, {Padilla}, {Sawicki}, {Seo}, {Singh}, {Song}, \& {Troncoso Iribarren}}]{Nicole24}
{Firestone}, N.~M., {Gawiser}, E., {Ramakrishnan}, V., {et~al.} 2024, \apj, 974, 217, \dodoi{10.3847/1538-4357/ad71c9}

\bibitem[{Firestone {et~al.}(2025)Firestone, Gawiser, Iyer, Lee, Ramakrishnan, Valdes, Park, Yang, Alavi, Ciardullo, {et~al.}}]{Nicole25}
Firestone, N.~M., Gawiser, E., Iyer, K.~G., {et~al.} 2025, ApJ in press, arXiv:2501.08568

\bibitem[{{Gawiser} {et~al.}(2007){Gawiser}, {Francke}, {Lai}, {Schawinski}, {Gronwall}, {Ciardullo}, {Quadri}, {Orsi}, {Barrientos}, {Blanc}, {Fazio}, {Feldmeier}, {Huang}, {Infante}, {Lira}, {Padilla}, {Taylor}, {Treister}, {Urry}, {van Dokkum}, \& {Virani}}]{Gawiser07}
{Gawiser}, E., {Francke}, H., {Lai}, K., {et~al.} 2007, \apj, 671, 278, \dodoi{10.1086/522955}

\bibitem[{Guaita {et~al.}(2010)Guaita, Gawiser, Padilla, Francke, Bond, Gronwall, Ciardullo, Feldmeier, Sinawa, Blanc, {et~al.}}]{Guaita10}
Guaita, L., Gawiser, E., Padilla, N., {et~al.} 2010, The Astrophysical Journal, 714, 255

\bibitem[{Hamilton(1998)}]{Hamilton}
Hamilton, A. J.~S. 1998, Linear Redshift Distortions: A Review (Springer Netherlands), 185–275, \dodoi{10.1007/978-94-011-4960-0_17}

\bibitem[{{Hao} {et~al.}(2018){Hao}, {Huang}, {Xia}, {Zheng}, {Jiang}, \& {Li}}]{Hao}
{Hao}, C.-N., {Huang}, J.-S., {Xia}, X., {et~al.} 2018, \apj, 864, 145, \dodoi{10.3847/1538-4357/aad80b}

\bibitem[{Harikane {et~al.}(2016)Harikane, Ouchi, Ono, More, Shimasaku, Saito, Lin, Coupon, Toshikawa, Komiyama, \& Miyazaki}]{Harikane2016}
Harikane, Y., Ouchi, M., Ono, Y., {et~al.} 2016, The Astrophysical Journal, 821, 123, \dodoi{10.3847/0004-637X/821/2/123}

\bibitem[{{Hong} {et~al.}(2019){Hong}, {Dey}, {Lee}, {Orsi}, {Gebhardt}, {Vogelsberger}, {Hernquist}, {Xue}, {Jung}, {Finklestein}, {Tuttle}, \& {Boylan-Kolchin}}]{Hong_2019}
{Hong}, S., {Dey}, A., {Lee}, K.-S., {et~al.} 2019, \mnras, 483, 3950, \dodoi{10.1093/mnras/sty3219}

\bibitem[{{Hu} \& {McMahon}(1996)}]{Hu_1996}
{Hu}, E.~M., \& {McMahon}, R.~G. 1996, \nat, 382, 231, \dodoi{10.1038/382231a0}

\bibitem[{{Jarvis} {et~al.}(2004){Jarvis}, {Bernstein}, \& {Jain}}]{Jarvis_2004}
{Jarvis}, M., {Bernstein}, G., \& {Jain}, B. 2004, \mnras, 352, 338, \dodoi{10.1111/j.1365-2966.2004.07926.x}

\bibitem[{{Kova{\v{c}}} {et~al.}(2007){Kova{\v{c}}}, {Somerville}, {Rhoads}, {Malhotra}, \& {Wang}}]{Kova__2007}
{Kova{\v{c}}}, K., {Somerville}, R.~S., {Rhoads}, J.~E., {Malhotra}, S., \& {Wang}, J. 2007, \apj, 668, 15, \dodoi{10.1086/520668}

\bibitem[{{Kusakabe} {et~al.}(2018){Kusakabe}, {Shimasaku}, {Ouchi}, {Nakajima}, {Goto}, {Hashimoto}, {Konno}, {Harikane}, {Silverman}, \& {Capak}}]{Kasukabe}
{Kusakabe}, H., {Shimasaku}, K., {Ouchi}, M., {et~al.} 2018, \pasj, 70, 4, \dodoi{10.1093/pasj/psx148}

\bibitem[{Kusakabe {et~al.}(2020)Kusakabe, Blaizot, Garel, Verhamme, Bacon, Richard, Hashimoto, Inami, Conseil, Guiderdoni, Drake, Christian~Herenz, Schaye, Oesch, Matthee, Anna~Marino, Borello~Schmidt, Pelló, Maseda, Leclercq, Kerutt, \& Mahler}]{Kusakabe2020}
Kusakabe, H., Blaizot, J., Garel, T., {et~al.} 2020, Astronomy \& Astrophysics, 638, A12, \dodoi{10.1051/0004-6361/201937340}

\bibitem[{{Landy} \& {Szalay}(1993)}]{Landy_1993}
{Landy}, S.~D., \& {Szalay}, A.~S. 1993, \apj, 412, 64, \dodoi{10.1086/172900}

\bibitem[{{Lee} {et~al.}(2009){Lee}, {Giavalisco}, {Conroy}, {Wechsler}, {Ferguson}, {Somerville}, {Dickinson}, \& {Urry}}]{Lee2009}
{Lee}, K.-S., {Giavalisco}, M., {Conroy}, C., {et~al.} 2009, \apj, 695, 368, \dodoi{10.1088/0004-637X/695/1/368}

\bibitem[{{Lee} {et~al.}(2024){Lee}, {Gawiser}, {Park}, {Yang}, {Valdes}, {Lang}, {Ramakrishnan}, {Moon}, {Firestone}, {Appleby}, {Artale}, {Andrews}, {Bauer}, {Benda}, {Broussard}, {Chiang}, {Ciardullo}, {Dey}, {Farooq}, {Gronwall}, {Guaita}, {Huang}, {Hwang}, {Im}, {Jeong}, {Karthikeyan}, {Kim}, {Kim}, {Kumar}, {Nagaraj}, {Nantais}, {Padilla}, {Park}, {Pope}, {Popescu}, {Schlegel}, {Seo}, {Singh}, {Song}, {Troncoso}, {Vivas}, {Zabludoff}, \& {Zenteno}}]{ODIN}
{Lee}, K.-S., {Gawiser}, E., {Park}, C., {et~al.} 2024, \apj, 962, 36, \dodoi{10.3847/1538-4357/ad165e}

\bibitem[{{Limber}(1953)}]{Limber}
{Limber}, D.~N. 1953, The Astrophysical Journal, 117, 134, \dodoi{10.1086/145711}

\bibitem[{Louis {et~al.}(2025)Louis, Posta, Atkins, Jense, Abril-Cabezas, Addison, Ade, Aiola, Alford, Alonso, Amiri, An, Austermann, Barbavara, Battaglia, Battistelli, Beall, Bean, Beheshti, Beringue, Bhandarkar, Biermann, Bolliet, Bond, Calabrese, Capalbo, Carrero, Chen, Chesmore, mei Cho, Choi, Clark, Cothard, Coughlin, Coulton, Crichton, Crowley, Darwish, Devlin, Dicker, Duell, Duff, Duivenvoorden, Dunkley, Dunner, Villagra, Fankhanel, Farren, Ferraro, Foster, Freundt, Fuzia, Gallardo, Garrido, Gerbino, Giardiello, Gill, Givans, Gluscevic, Goldstein, Golec, Gong, Guan, Halpern, Harrison, Hasselfield, Healy, Henderson, Hensley, Hervías-Caimapo, Hill, Hilton, Hilton, Hincks, Hložek, Ho, Hood, Hornecker, Huber, Hubmayr, Huffenberger, Hughes, Ikape, Irwin, Isopi, Joshi, Keller, Kim, Knowles, Koopman, Kosowsky, Kramer, Kusiak, Lague, Lakey, Lee, Li, Li, Limon, Lokken, Lungu, MacCrann, MacInnis, Madhavacheril, Maldonado, Maldonado, Mallaby-Kay, Marques, van Marrewijk, McCarthy, McMahon, Mehta, Menanteau,
  Moodley, Morris, Mroczkowski, Naess, Namikawa, Nati, Nerval, Newburgh, Nicola, Niemack, Nolta, Orlowski-Scherer, Pagano, Page, Pandey, Partridge, Sarmiento, Prince, Puddu, Qu, Ragavan, Guachalla, Rogers, Rojas, Sakuma, Schaan, Schmitt, Sehgal, Shaikh, Sherwin, Sierra, Sievers, Sifón, Simon, Sonka, Spergel, Staggs, Storer, Surrao, Switzer, Tampier, Thornton, Trac, Tucker, Ullom, Vale, Engelen, Lanen, Vargas, Vavagiakis, Wagoner, Wang, Wenzl, Wollack, \& Zheng}]{act}
Louis, T., Posta, A.~L., Atkins, Z., {et~al.} 2025, The Atacama Cosmology Telescope: DR6 Power Spectra, Likelihoods and $Λ$CDM Parameters.
\newblock \doarXiv{2503.14452}

\bibitem[{{Momose} {et~al.}(2021){Momose}, {Shimasaku}, {Nagamine}, {Shimizu}, {Kashikawa}, {Ando}, \& {Kusakabe}}]{Momose}
{Momose}, R., {Shimasaku}, K., {Nagamine}, K., {et~al.} 2021, \apjl, 912, L24, \dodoi{10.3847/2041-8213/abf04c}

\bibitem[{{Ouchi} {et~al.}(2010){Ouchi}, {Shimasaku}, {Furusawa}, {Saito}, {Yoshida}, {Akiyama}, {Ono}, {Yamada}, {Ota}, {Kashikawa}, {Iye}, {Kodama}, {Okamura}, {Simpson}, \& {Yoshida}}]{Ouchi10}
{Ouchi}, M., {Shimasaku}, K., {Furusawa}, H., {et~al.} 2010, \apj, 723, 869, \dodoi{10.1088/0004-637X/723/1/869}

\bibitem[{{Ouchi} {et~al.}(2018){Ouchi}, {Harikane}, {Shibuya}, {Shimasaku}, {Taniguchi}, {Konno}, {Kobayashi}, {Kajisawa}, {Nagao}, {Ono}, {Inoue}, {Umemura}, {Mori}, {Hasegawa}, {Higuchi}, {Komiyama}, {Matsuda}, {Nakajima}, {Saito}, \& {Wang}}]{Ouchi18}
{Ouchi}, M., {Harikane}, Y., {Shibuya}, T., {et~al.} 2018, \pasj, 70, S13, \dodoi{10.1093/pasj/psx074}

\bibitem[{Peebles(2020)}]{Peebles_2020}
Peebles, P. J.~E. 2020, The large-scale structure of the universe, Vol.~98 (Princeton University Press)

\bibitem[{{Ramakrishnan} {et~al.}(2023){Ramakrishnan}, {Moon}, {Im}, {Farooq}, {Lee}, {Gawiser}, {Yang}, {Park}, {Hwang}, {Valdes}, {Artale}, {Ciardullo}, {Dey}, {Gronwall}, {Guaita}, {Jeong}, {Padilla}, {Singh}, \& {Zabludoff}}]{ramakrishnan}
{Ramakrishnan}, V., {Moon}, B., {Im}, S.~H., {et~al.} 2023, \apj, 951, 119, \dodoi{10.3847/1538-4357/acd341}

\bibitem[{{Ramakrishnan} {et~al.}(2024){Ramakrishnan}, {Lee}, {Artale}, {Gawiser}, {Yang}, {Park}, {Chiang}, {Ciardullo}, {Dey}, {Gronwall}, {Guaita}, {Hwang}, {Im}, {Jeong}, {Kim}, {Kumar}, {Lee}, {Lee}, {Moon}, {Padilla}, {Pope}, {Popescu}, {Singh}, {Song}, {Troncoso}, {Valdes}, \& {Zabludoff}}]{Ramakrishnan24}
{Ramakrishnan}, V., {Lee}, K.-S., {Artale}, M.~C., {et~al.} 2024, \apj, 977, 119, \dodoi{10.3847/1538-4357/ad83cb}

\bibitem[{{Rhoads} {et~al.}(2001){Rhoads}, {Malhotra}, {Dey}, {Stern}, {Spinrad}, \& {Jannuzi}}]{LALA}
{Rhoads}, J.~E., {Malhotra}, S., {Dey}, A., {et~al.} 2001, in American Astronomical Society Meeting Abstracts, Vol. 198, American Astronomical Society Meeting Abstracts \#198, 54.09

\bibitem[{{Sawicki} {et~al.}(2019){Sawicki}, {Arnouts}, {Huang}, {Coupon}, {Golob}, {Gwyn}, {Foucaud}, {Moutard}, {Iwata}, {Liu}, {Chen}, {Desprez}, {Harikane}, {Ono}, {Strauss}, {Tanaka}, {Thibert}, {Balogh}, {Bundy}, {Chapman}, {Gunn}, {Hsieh}, {Ilbert}, {Jing}, {LeF{\`e}vre}, {Li}, {Matsuda}, {Miyazaki}, {Nagao}, {Nishizawa}, {Ouchi}, {Shimasaku}, {Silverman}, {de la Torre}, {Tresse}, {Wang}, {Willott}, {Yamada}, {Yang}, \& {Yee}}]{CFHT}
{Sawicki}, M., {Arnouts}, S., {Huang}, J., {et~al.} 2019, \mnras, 489, 5202, \dodoi{10.1093/mnras/stz2522}

\bibitem[{{Schechter}(1976)}]{Schechter}
{Schechter}, P. 1976, \apj, 203, 297, \dodoi{10.1086/154079}

\bibitem[{{Shapley} {et~al.}(2003){Shapley}, {Steidel}, {Pettini}, \& {Adelberger}}]{Shapley}
{Shapley}, A.~E., {Steidel}, C.~C., {Pettini}, M., \& {Adelberger}, K.~L. 2003, \apj, 588, 65, \dodoi{10.1086/373922}

\bibitem[{{Stark} {et~al.}(2010){Stark}, {Ellis}, {Chiu}, {Ouchi}, \& {Bunker}}]{Stark}
{Stark}, D.~P., {Ellis}, R.~S., {Chiu}, K., {Ouchi}, M., \& {Bunker}, A. 2010, \mnras, 408, 1628, \dodoi{10.1111/j.1365-2966.2010.17227.x}

\bibitem[{Tinker {et~al.}(2010)Tinker, Robertson, Kravtsov, Klypin, Warren, Yepes, \& Gottlöber}]{Tinker2010}
Tinker, J.~L., Robertson, B.~E., Kravtsov, A.~V., {et~al.} 2010, The Astrophysical Journal, 724, 878, \dodoi{10.1088/0004-637x/724/2/878}

\bibitem[{{Umeda} {et~al.}(2024){Umeda}, {Ouchi}, {Kikuta}, {Harikane}, {Ono}, {Shibuya}, {Inoue}, {Shimasaku}, {Liang}, {Matsumoto}, {Saito}, {Kusakabe}, {Kageura}, \& {Nakane}}]{umeda}
{Umeda}, H., {Ouchi}, M., {Kikuta}, S., {et~al.} 2024, arXiv e-prints, arXiv:2411.15495, \dodoi{10.48550/arXiv.2411.15495}

\bibitem[{{Watkins} {et~al.}(2024){Watkins}, {van der Marel}, \& {Bennet}}]{LMC}
{Watkins}, L.~L., {van der Marel}, R.~P., \& {Bennet}, P. 2024, \apj, 963, 84, \dodoi{10.3847/1538-4357/ad1f58}

\bibitem[{White {et~al.}(2024)White, Raichoor, Dey, Garrison, Gawiser, Lang, soo Lee, Myers, Schlegel, Valdes, Aguilar, Ahlen, Brooks, Chaussidon, Claybaugh, Dawson, de~la Macorra, Dey, Doel, Fanning, Font-Ribera, Forero-Romero, Gontcho, Gutierrez, Guy, Honscheid, Kirkby, Kremin, Landriau, Guillou, Levi, Magneville, Manera, Martini, Meisner, Miquel, Moon, Newman, Niz, Palanque-Delabrouille, Park, Percival, Prada, Rossi, Ruhlmann-Kleider, Sanchez, Schlafly, Schubnell, Seo, Sprayberry, Tarlé, Weaver, Yang, Yèche, \& Zou}]{White}
White, M., Raichoor, A., Dey, A., {et~al.} 2024, Journal of Cosmology and Astroparticle Physics, 2024, 020, \dodoi{10.1088/1475-7516/2024/08/020}

\bibitem[{{White} \& {Rees}(1978)}]{WhiteRees}
{White}, S.~D.~M., \& {Rees}, M.~J. 1978, \mnras, 183, 341, \dodoi{10.1093/mnras/183.3.341}

\end{thebibliography}
\bibliographystyle{aasjournal}

\end{document}